\documentclass{aipproc}

\usepackage{graphicx}

\layoutstyle{6x9}

\begin{document}

\title{Simulating the scalar glueball on the lattice}
\author{Colin Morningstar}{address={
Department of Physics, Carnegie Mellon University,
Pittsburgh, PA 15213-3890}}
\author{Mike Peardon}{address={School of Mathematics, Trinity College, Dublin
2, Ireland.}}

\begin{abstract}
Techniques for efficient computation of the scalar glueball mass on the lattice
are described. Directions and physics goals of proposed future calculations
will be outlined. 
\end{abstract}
\maketitle 

\section{Introduction}

Along with the mesons of the quark model, QCD allows for the existence of
bound-states of gluons, the glueballs. 
The scalar glueball is the lightest state in the spectrum of strongly
interacting gluons (described by the $SU(3)$ Yang-Mills theory) and in this
consistent, confining quantum field theory, it is stable. The world of gluons
alone 
thus provides an attractive starting point for investigations of glueballs.
Confinement is not described at all within QCD perturbation theory and so
computing properties of the glueballs requires a non-perturbative technique for
studying strong interactions.  
As a result, no
unambiguous identification of the scalar glueball has been made, although 
there is broad agreement that an extra state in the family of scalar 
resonances, unaccounted for by the quark model, exists between 1.5 and 
1.8 GeV. 
In QCD, the glueball states will mix with the isoscalar resonances of the 
quark model, and will also decay strongly so 
to claim a complete understanding of the scalar mesons below 2 GeV 
therefore requires reliable knowledge of the masses and other properties of 
glueballs. 

In order to disentangle the complex picture presented by experiments much
more detailed and precise theoretical prediction from QCD must be made. At 
present, the only practical {\it ab initio} method for computing the 
non-perturbative properties of strongly interacting field theories such as QCD
is through Monte Carlo simulation of the lattice regularisation of the theory.
The glueballs of $SU(3)$ Yang-Mills theory have been studied
extensively in lattice simulations \cite{Michael:1989jr, Bali:1993fb, 
Morningstar:1999rf}, and this spectrum is becoming increasingly
accurately determined. This is the first step in understanding the
appearance of glueballs among the light scalar resonances of QCD. In spite of 
the limitation of these calculations, their output is proving to be very 
valuable in the construction and validation of phenomenological models.

In this article, we describe the state-of-the-art calculation of the
glueball spectrum of the Yang-Mills theory. Investigations of further 
properties of these fascinating states, including their behaviour in QCD (with 
the dynamics of quark fields included) are then outlined.

\section{Lattice QCD}

Lattice QCD solves two key problems arising from strongly interacting gauge
theories at a stroke; it provides both a non-perturbative regularisation of QCD
and a means by which observables in the theory can be predicted by computer.
This inherently gauge invariant regularisation is manifest through 
a direct cut-off of momenta at the Brillouin zone boundaries of the lattice. 

To begin, the path-integral formulation of the theory in Euclidean space is
taken. The four dimensions of space and time are discretised with a regular
lattice of points separated by a spacing, $a$. In the gauge-invariant 
formulation defined by Wilson nearly 30 years ago \cite{Wilson:1974sk}, the quark fields of QCD are defined
at sites of the lattice, while the gluonic degrees of freedom are represented
by parallel transporters (usually in the fundamental representation of the
gauge group) connecting nearest-neighbouring sites. A definition of the
Yang-Mills action is constructed from the trace of path-ordered products of the
link variables around small loops on the lattice. The smallest non-trivial loop
on the lattice circumnavigates a square of side-length $a$, usually called a
plaquette.  Similarly, a number of different definitions of the quark bilinear 
action, coupled to the gluons, can be made. 

Path integral expectation values can then be estimated by numerical methods.
If the theory is considered for space and time inside a finite-sized box, the
path integral at non-zero lattice spacing is reduced to a very-high-dimensional
ordinary integral. This type of problem can normally only be addressed by
Monte Carlo methods, where a stochastic sampling of points inside the phase
space to be integrated over is made.

\subsection{Stochastic estimation methods for quantum field theory}

After Wick rotation, the path integral for field theories such as QCD (at least 
at zero chemical potential) can be regarded as a statistical mechanical
partition function. The Boltzmann weight for a particular field configuration
is real and positive definite, and so can be interpreted as a statistical 
probability of the configuration appearing. In these circumstances, the 
natural Monte Carlo method to employ is importance sampling. An update 
algorithm is defined which generates configurations of the gluon fields, $U$ 
with probability density given by the Boltzmann weight, $\exp\{ -S(U)\}$. The 
method is employed on the computer to generate an ensemble of gluon field 
configurations. Expectation values are then estimated from ensemble averages. 

To compute the physical properties of the theory appropriate observation
functions on the underlying fields are defined with the required quantum 
numbers, and these observables are measured on all members of the stochastic 
ensemble.  Careful analysis allows a reliable determination of statistical 
uncertainties.

\subsection{Controlling simulation artefacts}

In order to make contact with the continuum field theory, a number of artefacts 
of the method must be controlled. The most obvious of these is the need to use
a finite-sized box for numerical work. In practical simulations, boxes with
side lengths of the order of 2 fm are feasible. 

In many cases, the qualitative behaviour of states in increasingly 
large-but-finite volumes can be predicted. Simulations are run for a range of 
box sizes, and the data matched to this predicted form to allow safe 
extrapolation to the infinite volume limit. In most studies of glueballs these 
effects are now extremely small and do not constitute the dominant systematic 
error. 

More difficult is the need to extrapolate data to the limit of zero lattice
spacing (usually called the continuum limit). Again, the most natural procedure
is to compute the physics of interest on a set of lattices with different grid
spacings, and extrapolate. In many cases, the expected leading-order behaviour
of these finite lattice spacing effects can be predicted, and this can be used 
to control data fitting. The problem with this direct approach is that the cost 
of computer simulations rises rapidly as the lattice spacing is diminished, at
least in proportion to $a^{-4}$ and generally worse than this. Since the
continuum limit exists at a critical point corresponding to a second order
phase transition, where the correlation length of the system diverges in units
of the lattice spacing, critical slowing down makes this cost higher. It is
clear that the cheapest computer simulations are those run on the coarsest
grids, which are furthest from the continuum. 

\subsection{Symanzik improvement}

The effects of a finite grid spacing can be understood in the language of
quantum field theory as arising from irrelevant operators appearing in the
lattice action. These operators are of higher dimension than the relevant 
operator of the continuum theory, Tr $F_{\mu\nu} F_{\mu\nu}$ and are 
multiplied in the action by dimensionful couplings proportional to (positive)
powers of the lattice spacing. For the simplest description of the gauge
action, proposed by Wilson and consisting of the trace of the plaquette, the
leading irrelevant operator appears at ${\cal O}(a^2)$. Odd powers are
prohibited by exact parity symmetry, which is preserved by the discretisation. 

Universality suggests that there is a broad class of lattice
representations of the continuum action, whose members all have the identical
continuum limit: QCD. This implies that different lattice actions can be 
engineered which have smaller contributions from higher-dimensional operators. 
For asymptotically free theories, like QCD and the $SU(3)$ Yang-Mills theory of
gluons, these discretisations of the action can be designed within the 
framework of perturbation theory. This concept is termed Symanzik improvement
\cite{Symanzik:1983dc}. 

The idea has been widely exploited in lattice calculations, both in
discretisations of gluon fields as well as the quark fields. For the gluon
action, taking appropriate linear combinations of traces of larger closed
loops, such as the $2\times 1$ rectangle means terms at ${\cal O}(a^2)$ can be
eliminated \cite{Luscher:1985zq, Lepage:1996ph}. This allows lattice 
investigations to be carried out with grid spacings as coarse as 
0.25 fm, while remaining sufficiently close to the continuum limit to 
explore the physics reliably. 

\subsection{The anisotropic lattice}

At the point in the calculation where measurements are made, these coarse 
lattices present new disadvantages. The mass of a state is extracted from the 
decay of a two-point correlation function in Euclidean space
\begin{equation}
    C(t) = \langle \Phi(t) \Phi(0) \rangle \propto e^{-M t}
\end{equation}
where $\Phi(t)$ is a lattice operator that creates or annihiliates the state 
of mass $M$ at time $t$. This correlator falls exponentially as the
two operators are moved further apart, and for a heavy state (such as the
glueballs) the fall-off is rapid. Also, since the correlation function is being
estimated in a Monte Carlo simulation, the statistical variance of the
operators determines the accuracy to which these correlations can be measured.
For the glueballs, these operators are the traces of closed Wilson loops on the
underlying gauge fields and have rather large vacuum fluctuations, so a
critical signal-to-noise problem arises.  Since the operators can only be
measured on points of the grid, little information can be extracted from coarse
lattice measurements before the signal is lost in the noisy vacuum.

A solution that combines the economy of the coarse lattice with the good
resolution of a fine grid is to use an anisotropic lattice, where the temporal
lattice spacing is made small, while the three spatial dimensions are
discretised more coarsely. Note that time-like correlation functions are used
in extracting the masses of states. This method is particularly efficient for
glueball simulations \cite{Morningstar:1997ff}. The two natural length scales 
in the problem, the mass and size of the glueball set the optimal temporal and
spatial lattice spacings for numerical investigation.

\begin{center}
\begin{figure}
\includegraphics[width=12cm]{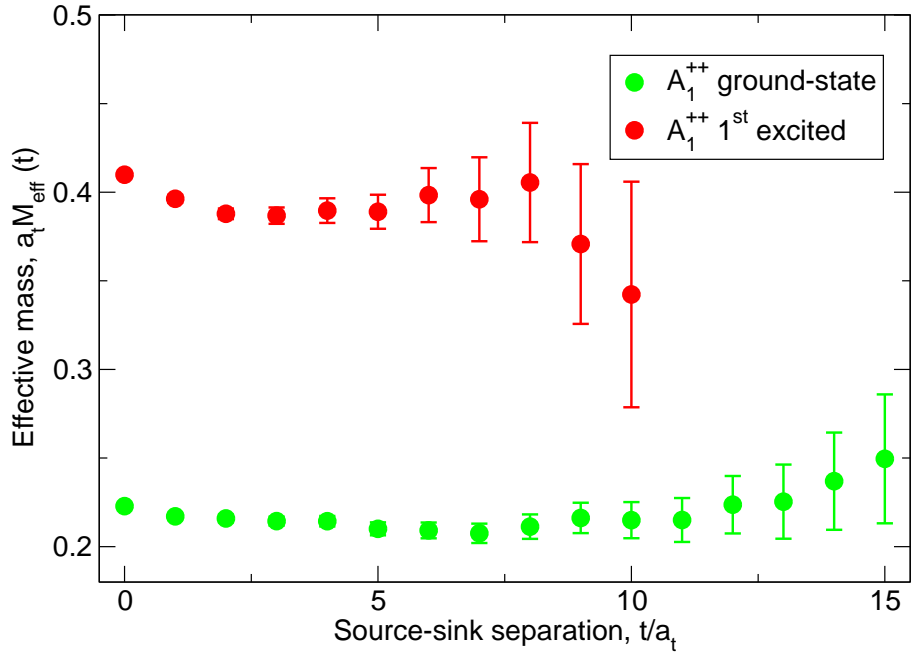}
\caption{The effective mass plot of the scalar glueball channel, including data
for the first excited state.\label{fig:meff}}
\end{figure}
\end{center}

Fig.~\ref{fig:meff} describes the output from just such an investigation. The 
figure shows the effective mass,
\begin{equation}
   a_t M_{\rm eff} = \frac{\ln\; C(t)}{\ln\; C(t+1)}
\end{equation}
for the scalar glueball measured on a highly anisotropic lattice, where the
ratio of scales is $a_s/a_t=6$. The operator, $\Phi$ is constructed from a
large basis set of the trace of closed Wilson loops built from smoothed link
variables. A single operator for each state is made by taking linear
combinations from the basis set of operators, with this combination chosen to
optimise the overlap with the state of interest. The graph shows data for the 
ground-state scalar and the first excited state with the same quantum numbers.
It illustrates the method is capable of precise determinations of the energies 
of both these levels. In the simulation presented in Fig.~\ref{fig:meff} these
energies are measured to $1\%$ statistical precision.

In Fig.~\ref{fig:dip}, the dependence on the lattice cut-off of the scalar
glueball mass in units of a physical scale $r_0$, derived from the static
potential \cite{Sommer:1994ce}, is presented. The data points
represented with crosses are from a range of simulations performed using
the Wilson plaquette action \cite{Michael:1989jr, Bali:1993fb}. The circles 
are from simulations with an 
anisotropic lattice action, improved according to the Symanzik programme with 
parameters determined from tree-level perturbation theory after the tadpole 
graphs in the lattice weak-coupling expansion have been resummed 
\cite{Morningstar:1997ff, Morningstar:1999rf}. The quality 
of a lattice action should be judged by how strongly the mass (in physical 
units) depends on the cut-off; ``better'' actions should show a weaker 
dependence on the lattice spacing. By this assessment, the Symanzik improved 
action is superior to the simplest action, however a strong dependence on the 
cut-off still remains. The scalar glueball mass (in units of $r_0^{-1}$) falls 
as the lattice spacing is increased to reach a minimum from which it begins to 
rise again.  We term this effect the ``scalar dip.''

\subsection{Curing the ``scalar dip''}

The lattice theory is not necessarily QCD; the physics of this theory is only 
recovered once the discretised version is ``close'' to the critical point on 
which QCD exists. If the space of lattice theories contains other critical 
points, simulations performed near to these points will be probing the 
properties of other continuum quantum field theories. Remarkably, this 
apparently abstract problem in non-perturbative renormalisation seems to arise 
in the simulation of $SU(3)$ Yang-Mills. 

It has been recognised for some time that there is a line of ``bulk'' 
first-order phase transitions in the two-dimensional plane of lattice theories 
in which both a coupling to the fundamental and adjoint representations of the 
link variables is made \cite{Heller:1995bz}. This line ends in a critical 
point at which some
unknown continuum theory resides. If the physics of the lattice theory close
to this critical point is investigated, it will be predominantly governed by
the properties of this other theory. One suggestion is that the continuum
theory at this critical point is a free scalar one, and if this was the case,
the mass of the scalar particle in the lattice theory would be artificially
light in comparison to the higher spin states. Precisely this artefact is
observed. 

\begin{center}
\begin{figure}
\includegraphics[width=12cm]{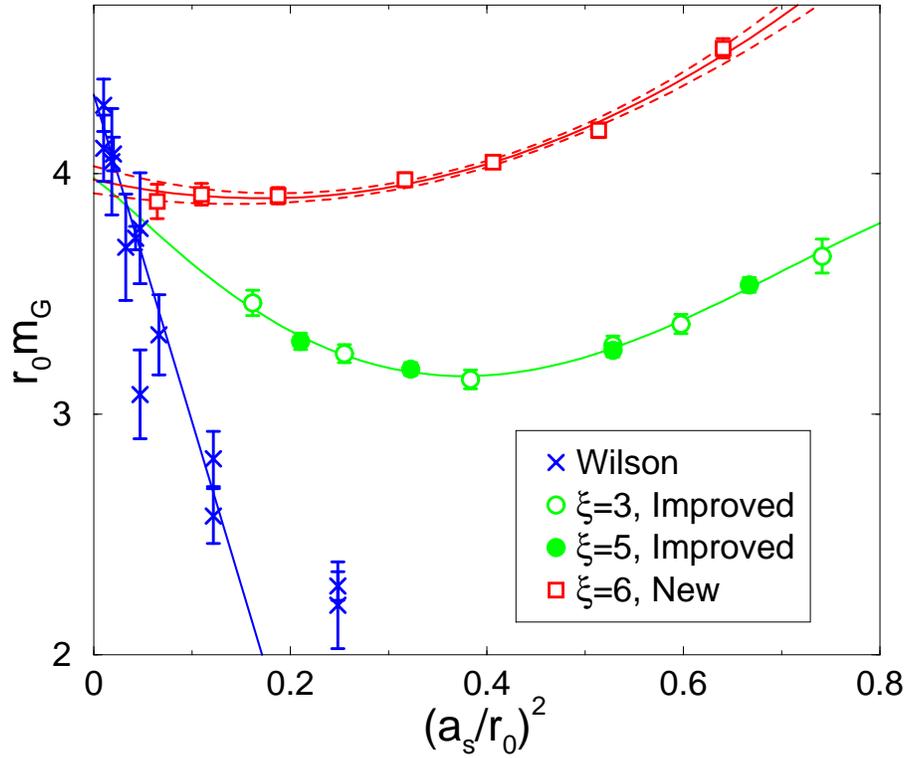}
\caption{Extrapolating simulation data at finite lattice spacing to the
continuum limit.\label{fig:dip}}
\end{figure}
\end{center}

The data represented by squares in Fig.~\ref{fig:dip} are from a new
discretisation of the gluon action on an anisotropic lattice. In this
prescription, the lattice action includes terms that trace over the 
link variables in the adjoint representation, as well as the fundamental. 
The links stored on the computer are in the fundamental representation, but the
trace in the adjoint can be computed using the identity
\begin{equation}
  \mbox{Tr }A(g) = \mbox{Tr }U(g) \;\mbox{Tr }U^\dagger(g) - 1,
\end{equation}
with $A(g)$ the adjoint representation of an element, $g \in SU(3)$ and $U(g)$
its corresponding fundamental representation. 

These simulation results are extremely encouraging and clearly show very weak 
lattice spacing dependence out to coarse spatial lattices ($a\approx 0.25$ fm).
A very reliable extrabolation to the continuum limit can be made.

\subsection{Spin on the lattice}

In a scattering experiment, the spin of resonances is determined from a
partial wave analysis. Revealing the spin of states in a lattice
calculation requires similar care.
Putting quarks and gluons onto a lattice breaks the continuum rotation group 
$SO(3)$ to the discrete cubic point group, $O_h$. As a result, states are no
longer classified according to a spin quantum number which describes the
irreducible representation (irrep) of $SO(3)$ they transform under.
They have instead one of five labels corresponding to the five irreps of $O_h$:
$A_1,A_2,E,T_1$ and $T_2$. 
Table~\ref{tab:subduce}
describes how the representation of $O_h$ subduced from spin irreps of 
$SO(3)$ subsequently decompose into irreps of $O_h$.
\begin{table}[h]
\begin{tabular}{c|ccccc}
  J & $A_1$ & $A_2$ & $E$ & $T_1$ & $T_2$ \\
  \hline
  0 &   1   &       &     &       &       \\
  1 &       &       &     &   1   &       \\
  2 &       &       &  1  &       &  1    \\
  3 &       &   1   &     &   1   &  1    \\
  4 &   1   &       &  1  &   1   &  1    \\
  \hline
\end{tabular}
\caption{The irreducible content of the representations of $O_h$ subduced from
the spin irreps of $SO(3)$}\label{tab:subduce}
\end{table}
To identify the spin of the state in the continuum,
degeneracies across lattice states must be found and compared to the table. For
example, tagging a spin 2 state on the lattice requires finding degenerate
states in the $E$ and $T_2$ channels (and no others) in the continuum limit. 
For the
scalar state, this procedure is quite straightforward; a state in the trivial
representation, $A_1$, which is not degenerate with any other state in the
spectrum can be uniquely identified with a continuum $J=0$ scalar. This spin
analysis was carried out in Ref.~\cite{Morningstar:1999rf} for the glueball 
spectrum, and the spins of many continuum states were identified. 

\section{The $SU(3)$ Yang-Mills spectrum}

\begin{center}
\begin{figure}[h!]
\includegraphics[width=12cm]{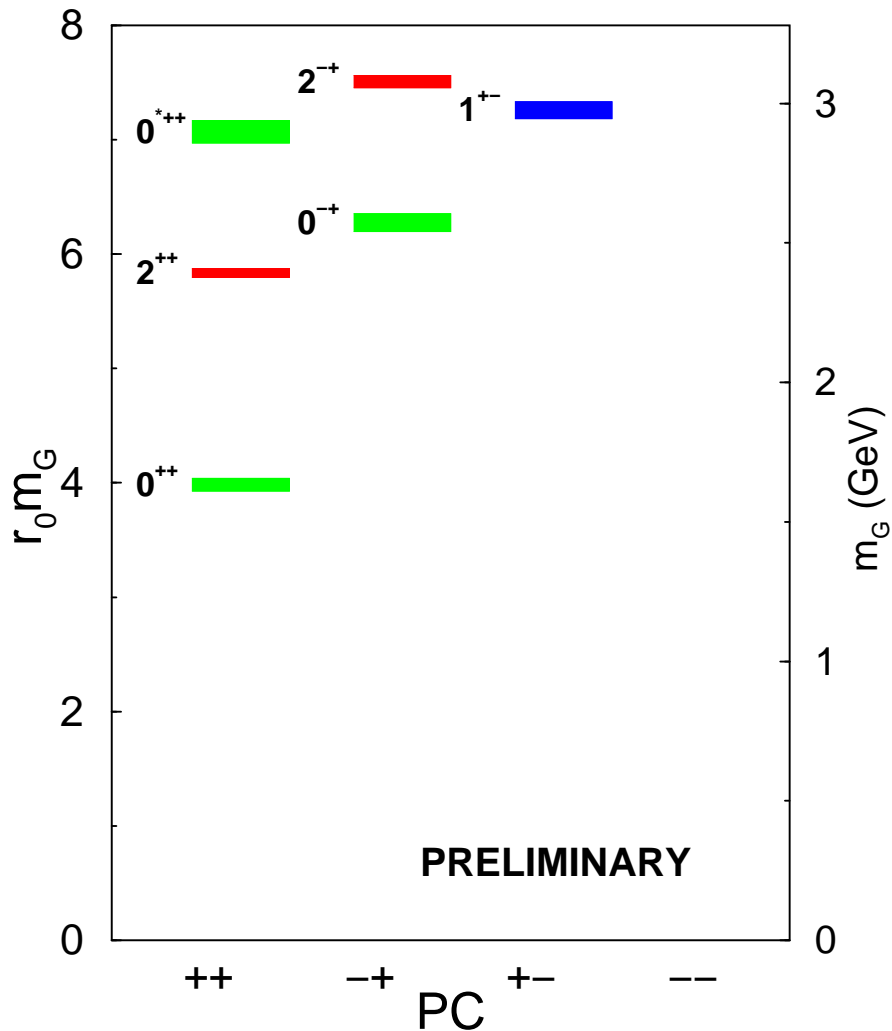}
\caption{The lightest six states in the spectrum of the $SU(3)$ Yang-Mills
theory.\label{fig:spectrum}}
\end{figure}
\end{center}

Following the procedures outlined in the previous sections, the simulations of 
the Yang-Mills theory of strongly interacting gluons (without quarks) have 
reached high levels of precision. Fig.~\ref{fig:spectrum} shows the lightest
six states in this spectrum of gluon bound-states. Note that both the 
ground-state and first-excited state in the scalar channel has been determined 
more reliably after removing the artefacts of the scalar dip and performing a
continuum extrapolation. The masses of the scalar glueballs are consistent
within errors with earlier determinations. 

\section{Current and future directions}

While the simulation of the Yang-Mills theory is now well understood, the
inclusion of quantum fluctuations of quark fields is significantly more
difficult. Since the Grassmann algebra of fermions in a path integral can not 
be handled directly by computers, the quark bilinear in the QCD action must be 
integrated out analytically, leaving a non-local effective action in the 
gluons alone. The non-local nature of these interactions means the simpler 
Monte Carlo algorithms used to generate ensembles of gauge field configurations 
for the Yang-Mills theories become inefficient. More sophisticated techniques 
are used, but they add a large numerical overhead; QCD simulations including 
quark dynamics are a factor of 100-1000 times more expensive than those of the 
Yang-Mills theory. 

This issue of including the quark field dynamics remains a topic of a good deal
of research in the lattice community and optimal simulation strategies are
still actively under investigation. It seems likely that within the near
future, the physics of glueballs that can decay into and mix with quark mesons
will be investigated. Progress in this direction has begun by other lattice
groups \cite{Hart:2002sp} although lattice calculations involving unstable 
particles are in their infancy and remaind a challenging topic of research. 
Also, extending the anisotropic lattice technology, 
which proved so useful for scanning the pure gauge theory, to QCD simulations 
with dynamical quarks is under investigation.

\section{Conclusions}

Scalar mesons seem to constantly challenge theorists and phenomenologists by
turn. At this meeting we heard of the difficulties in finding a convincing
picture of the broad range experimental data and in this article, some of the 
quite distinct problems presented by the scalar glueball to lattice theorists 
have been outlined. Significant progress in this field is still being made and 
the Yang-Mills theory has now been mapped out. Full QCD dynamics, with the quark
fields playing their role, is a major challenge under investigation by
many members of the lattice community \cite{Lat03}. At the same time, more 
detailed properties of the gluonic theory are now being examined, including 
calculations of the chromoelectric and magnetic matrix elements 
\cite{Dong:Lat03} and the structure form factors. 

\bibliography{trinlat_bib/trinlat}

\end{document}